\documentclass[acmtog, balance = false]{acmart}
\acmSubmissionID{486}

\usepackage{booktabs} %

\copyrightyear{2024}
\acmYear{2024}
\setcopyright{acmlicensed}\acmConference[SIGGRAPH Conference Papers '24]{Special Interest Group on Computer Graphics and Interactive Techniques Conference Conference Papers '24}{July 27-August 1, 2024}{Denver, CO, USA}
\acmBooktitle{Special Interest Group on Computer Graphics and Interactive Techniques Conference Conference Papers '24 (SIGGRAPH Conference Papers '24), July 27-August 1, 2024, Denver, CO, USA}
\acmDOI{10.1145/3641519.3657433}
\acmISBN{979-8-4007-0525-0/24/07}



\usepackage{hyperref}
\usepackage{url}
\usepackage{wrapfig}
\usepackage{xcolor}
\usepackage{colortbl}
\usepackage{multirow}

\usepackage{wrapfig}
\usepackage{nicefrac}
\usepackage{mathtools}
\usepackage{graphicx}
\usepackage{booktabs} 
\usepackage{combelow} %
\usepackage{tabularx} %
\usepackage{colortbl} %

\usepackage{manfnt} %

\usepackage{caption}
\usepackage{subcaption}

\usepackage{amsmath}
\usepackage{amsfonts}
\usepackage{amsbsy}

\usepackage{cleveref}

\usepackage{listings}

\definecolor{codegreen}{rgb}{0,0.6,0}
\definecolor{codegray}{rgb}{0.5,0.5,0.5}
\definecolor{codepurple}{rgb}{0.58,0,0.82}
\definecolor{backcolour}{rgb}{0.95,0.95,0.92}

\lstdefinestyle{my_code_style}{
  backgroundcolor=\color{backcolour}, 
  commentstyle=\color{codegreen},
  keywordstyle=\color{magenta},
  numberstyle=\tiny\color{codegray},
  stringstyle=\color{codepurple},
  basicstyle=\ttfamily\footnotesize,
  breakatwhitespace=false,         
  breaklines=true,                 
  captionpos=b,                    
  keepspaces=true,                 
  numbers=left,                    
  numbersep=5pt,                  
  showspaces=false,                
  showstringspaces=false,
  showtabs=false,                  
  tabsize=2
}
\usepackage{xfrac}

\newcommand{\refequ}[1] {Eq.~\ref{equ:#1}}

\newcommand{\reffig}[1] {Fig.~\ref{fig:#1}}

\newcommand{\reftab}[1] {Table~\ref{tab:#1}}
\newcommand{\refsec}[1] {Sec.~\ref{sec:#1}}

\newcommand{\vecFont}[1]{\mathbf{#1}}

\def\vd{{\vecFont{d}}}

\def\vg{{\vecFont{g}}}

\def\vx{{\vecFont{x}}}

\newcommand{\matFont}[1]{\mathbf{#1}}

\def\mF{{\matFont{F}}}

\def\mH{{\matFont{H}}}

\def\mP{{\matFont{P}}}

\newcommand{\changed}[1]{{#1}}
\colorlet{RED}{red} %

\newlength\savedwidth

\lstdefinelanguage{Pseudocode}%
  {morekeywords={abstract,break,case,catch,const,continue,do,else,elseif,%
      end,export,false,for,function,immutable,import,importall,if,in,%
      macro,module,otherwise,quote,return,switch,true,try,type,typealias,%
      using,while,either,or,max,abs,foreach},%
   sensitive=true,%
   alsoother={$},%
   morecomment=[l]\#,%
   morecomment=[n]{\#=}{=\#},%
}[keywords,comments,strings]%

\definecolor{derekblue}{RGB}{144,187,195}
\definecolor{darkderekblue}{RGB}{86,130,140}
\definecolor{verylightgray}{RGB}{245,245,245}

\begin{document}

\newcommand{\abbrtitle}{Stabler Neo-Hookean Simulation}
\title[\abbrtitle]{\abbrtitle:\\
Absolute Eigenvalue Filtering for Projected Newton}

\author{Honglin Chen}
\affiliation{%
  \institution{Columbia University}
  \country{USA}
  }
\email{honglin.chen@columbia.edu}

\author{Hsueh-Ti Derek Liu}
  \affiliation{%
    \institution{Roblox Research \& University of British Columbia}
    \country{Canada}}
  \email{hsuehtil@gmail.com}

\author{David I.W. Levin}
\affiliation{%
  \institution{University of Toronto \& NVIDIA}
  \country{Canada}}
\email{diwlevin@cs.toronto.edu}

\author{Changxi Zheng}
  \affiliation{%
    \institution{Columbia University}
    \country{USA}}  
  \email{cxz@cs.columbia.edu}

\author{Alec Jacobson}
\affiliation{%
  \institution{University of Toronto \& Adobe Research}
  \country{Canada}}
\email{jacobson@cs.toronto.edu}

\begin{teaserfigure}
\centering
  \includegraphics[width=0.89\linewidth]{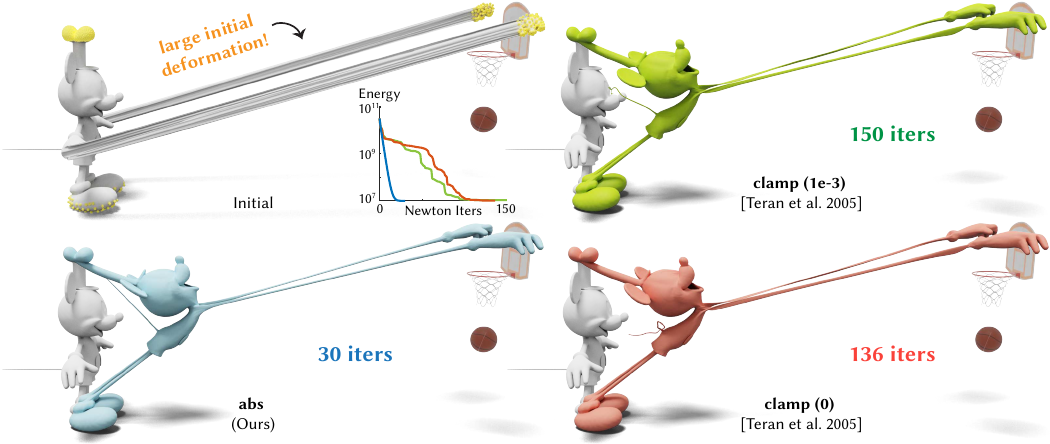}
  \caption{Our absolute eigenvalue projection scheme stabilizes the projected Newton optimization of stable Neo-Hookean energy under \emph{high Poisson's ratio} and \emph{large initial volume change}, and achieves a faster convergence rate than the traditional eigenvalue clamping scheme \cite{irving05}.
  Here we set the Poisson's ratio $\nu$ to be 0.495 and the fixed vertices are colored in yellow.
  }
  \label{fig:teaser}
\end{teaserfigure}

\begin{abstract}

Volume-preserving hyperelastic materials are widely used to model near-incompressible materials such as rubber and soft tissues.
However, the numerical simulation of volume-preserving hyperelastic materials is
  notoriously challenging within this regime due to the non-convexity of the
  energy function.
In this work, 
we identify the pitfalls of the popular eigenvalue clamping strategy 
for projecting Hessian matrices to positive semi-definiteness during 
Newton's method.
We introduce a novel eigenvalue filtering strategy for projected Newton's
method to stabilize the optimization of Neo-Hookean energy and other
volume-preserving variants under high Poisson's ratio (near 0.5) and large
initial volume change.
Our method only requires \emph{a single line} of code change in the existing
projected Newton framework, while achieving significant improvement in both
stability and convergence speed.
We demonstrate the effectiveness and efficiency of our eigenvalue projection
scheme on a variety of challenging examples and over different deformations on
a large dataset. 
\end{abstract}

\maketitle

\vspace{-1mm}
\section{Introduction}
Volume-preserving hyperelastic energies play a crucial role in accurately
capturing the near-incompressible nature of soft tissues and rubber-like
materials.
Among the common hyperelastic material models, Neo-Hookean model and its
variants are usually the de facto choice for modeling such materials with high
Poisson's ratios $\nu \in [0.45, 0.5)$ (see \reftab{materials_high_pr}).
Unfortunately, the numerical optimization of Neo-Hookean material is notoriously
challenging within this near-incompressible regime due to the non-convexity from
the volume-preserving term, especially in the presence of large volume change.
We will show that
existing projection approaches to ensure positive-definite of Hessian matrices during Newton's method
effectively assume the initial shape to have small volume
change, restraining the user from freely editing the initial deformation.

\setlength{\columnsep}{6pt}%
\begin{wraptable}[12]{r}{0.4\linewidth}
  \vspace{-12pt}
  \caption{Common materials with high Poisson's ratio (PR).}\label{tab:materials_high_pr}
  \vspace{-7pt}
    \begin{tabular}{ll}
    \hline
    Material       & PR \\ \hline
    Rubber         & 0.4999          \\
    Tongue         & 0.49            \\
    Soft palate    & 0.49            \\
    Fat            & 0.49            \\
    Skin           & 0.48            \\
    Muscle         & 0.47            \\
    Saturated clay & 0.4-0.49       \\
    \hline 
    \end{tabular}
    \vspace{-20pt}
\end{wraptable} 

We seek to stabilize and accelerate the optimization of Neo-Hookean energies
with high Poisson's ratios under the projected Newton framework in a
\emph{robust} and \emph{efficient} way.
For \emph{robustness}, we target the scenarios of large initial deformation and volume change, allowing the user to freely deform the initial shape without worrying about optimization blowing up.
For \emph{efficiency}, we aim to improve the convergence speed of projected
Newton method but still keep the per-step computation cost unchanged: filtering on the eigenvalues of per-element Hessian contributions \`a la \citet{irving05}.
Surprisingly, this leads to an extremely \emph{simple} and \emph{elegant} solution, requiring only one line of code change in the existing projected Newton framework without any additional parameter 
\textbf{(TL;DR):}

\begin{minipage}{0.41\linewidth}
  \label{code:change}
  \begin{lstlisting}[language=Pseudocode,mathescape=true]
  $\Lambda$,U=eig(Hi)
  Hi_proj=U*max($\Lambda$,0)*U'
\end{lstlisting}
\end{minipage}
\enspace $ \rightarrow $ \hfill
\begin{minipage}{0.41\linewidth}
\begin{lstlisting}[language=Pseudocode,mathescape=true]
  $\Lambda$,U=eig(Hi)
  Hi_proj=U*abs($\Lambda$)*U'
\end{lstlisting}
\end{minipage}

\begin{figure}[t]
  \centering
  \includegraphics[width=\linewidth]{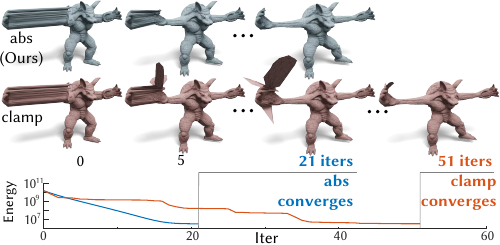}
  \caption{Our absolute eigenvalue projection strategy is robust to high Poisson's ratio and large volume change.} 
  \label{fig:armadillo_stretch}
  \vspace{-1mm}
\end{figure}

While our proposed change to code is small, the technical analysis behind it is non-trivial and interesting:
\begin{itemize} 
  \item 
      \changed{Our simple solution is supported by a thorough analysis. }
      We show that the high non-convexity of the Neo-Hookean energy stems from
      high Poisson's ratio and large volume change (\refsec{snh_eigenanalysis}).
    We analyze the behavior of projected Newton method under such
      scenarios
      and identify potential pitfalls of the traditional eigenvalue-clamping
      projection strategy \cite{irving05} (\refsec{pitfall_of_clamping}).
  \item In response, we propose a novel eigenvalue projection strategy for Newton's method to
    stabilize the optimization of Neo-Hookean-type energy under high Poisson's
    ratio and large initial volume change (\refsec{abs_projection}).
\end{itemize}
Our paper is a nice complement to the ``Stable Neo Hookean Flesh Simulation'' paper \cite{smith2018snh} which our title alludes to, improving the numerical stability by dissecting its nonconvexity in the projected Newton optimization.
We illustrate the effectiveness and efficiency of our eigenvalue projection
scheme on a wide range of challenging examples, including different
deformations, geometries, mesh resolutions, elastic energies, and physical
parameters. Through extensive experiments, we show that in the case of large
volume change, our method outperforms the state-of-the-art eigenvalue
projection strategy and other alternatives in terms of stability and
convergence speed, while still, on average, being comparable for simulations with moderate
volume changes. Our method is robust across different mesh resolutions and to
extreme volume change and inversion. It achieves a stable and fast convergence
rate in the presence of a high Poisson's ratio and large volume change.

\begin{figure}[t]
  \centering
  \includegraphics[width=\linewidth]{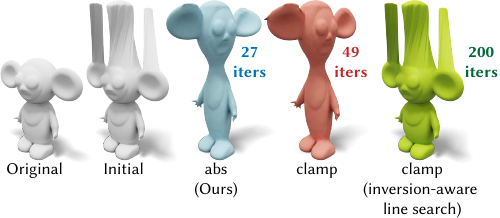}
  \caption{Inverted elements are common in the presence of large initial deformation, and thus adding an inversion-aware line search may make the optimization stall completely (right).
  Here we stretch and twist the topmost vertices of the doll by 90 degrees, resulting in 144 inverted elements in the initial deformation.}
  \label{fig:doll_twist_inv_ls}
  \vspace{-1mm}
\end{figure}

\section{Related Work}\label{sec:related}
While researchers have studied first-order methods for simulating volumetric
objects, our approach focuses on improving the robustness and efficiency of the
second-order methods, to which we limit our discussion.

\textit{Volume-preserving hyperelastic energies}
play a crucial role in capturing the volumetric behavior of deformable objects.
Many hyperelastic energies are modeled as functions of the deformation gradient $\mF$. 
As hyperelastic materials (e.g., rubber) resist volume changes, these energy
functions often contain a volume term, a function of the determinant of the
deformation gradient $\det(\mF)$, to penalize volume changes.
In practice, Neo-Hookean energy \cite{ogden1997non} and other volume-preserving
variants (e.g., volume-preserving ARAP \cite{lin2022isoARAP}) have been widely used 
for modeling materials with high Poisson's ratios.
Other alternatives such as St. Venant-Kirchhoff model~\cite{Picinbono2004stvk}
and co-rotational model~\cite{muller2002corotational} approximate or linearize
the volume term and thus compromise its volume-preserving property and visual
effects, particularly in the cases of large deformation and high Poisson's
ratios (see Sec.6.2 of \cite{kim22deformables}).

More recently, several works have focused on improving the stability and
computational efficiency of volume-preserving hyperelastic energies.
To improve robustness to mesh inversion and rest-state stability, \citet{smith2018snh}
proposed the stable Neo-Hookean energy
and further applied their analysis to stabilize other hyperelastic energies such as
Fung and Arruda-Boyce elasticity \cite{smith2018snh} and Mooney-Rivlin
elasticity \cite{kim22deformables}.
To improve computational efficiency, a series of works
\cite{smith2018snh,smith2019analytic,lin2022isoARAP} derive analytical
eigendecompositions for isotropic distortion energies to accelerate the
positive semi-definite (PSD) projection of per-element Hessian
contributions (following \changed{\cite{irving05}}).

\begin{figure}[t]
  \centering
  \includegraphics[width=\linewidth]{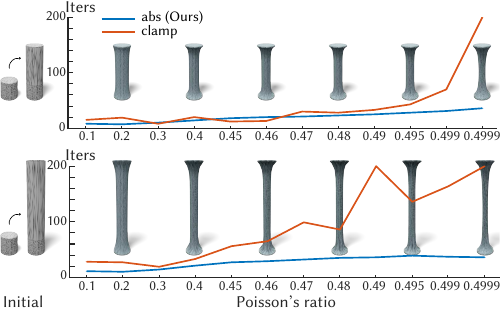}
  \caption{The speedup we obtained over the eigenvalue clamping strategy increases as the Poisson's ratio and the volume change increase.}
  \label{fig:different_pr}
  \vspace{-1mm}
\end{figure}

\paragraph{(Element-wise) projected Newton's method.}
When the Hessian matrix is not positive definite, steps in Newton's method
may not always find a direction leading to energy decrease~\cite{num_opt2006}. 
Several Hessian approximation strategies, a.k.a. projected Newton's methods,
have been adapted. %
Directly computing the eigenvalues and eigenvectors of a global Hessian matrix is
often too expensive.
For many energy functions, their Hessian matrices can be decomposed into a summation
over the sub-Hessian of each mesh element. 
Thereby, 
one can project each sub-Hessian to
the PSD cone by clamping negative eigenvalues to a small positive number (or
zero)~\cite{irving05} or adding a diagonal matrix~\cite{FuL16}. 
These approaches guarantee that the sub-Hessians are PSD and thus the resulting global Hessian is also PSD. 
While improving robustness, the projected Newton's method may suffer from
worse convergence rate compared to the classic Newton's method~\cite{longva2023pitfalls}. 
What is a better projection strategy remains an open question \cite{num_opt2006}. 

In this work, we analyze the limitations of existing
per-element projected Newton strategies, supported by extensive empirical
studies. 
In the optimization \changed{and machine learning} literature, \citet{gill1981practical} (Sec.4.4.2.1), \citet{Paternain2019nonconvex_newton} \changed{and \citet{dauphin2014saddle_free}} suggest projecting the negative eigenvalues of the \emph{global} Hessian to its absolute values.
We demonstrate that projecting the negative eigenvalues of the \emph{per-element} Hessian to its absolute values, similar to \cite{Paternain2019nonconvex_newton} \changed{and \cite{dauphin2014saddle_free}}, performs the best in hyperelastic simulations.

\begin{figure}[t]
  \centering
  \includegraphics[width=\linewidth]{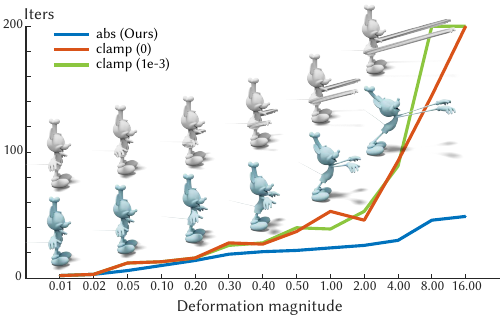}
  \caption{The speedup we obtained over the eigenvalue clamping strategy increases as the initial volume change increases, while still being comparable for moderate volume change cases.
  Here we visualize the initial deformation (white) and our final results (blue) under deformation magnitudes 0.01, 0.05, 0.20, 0.40, 1.00 and 2.00 (from left to right).}
  \label{fig:different_deformation_size}
  \vspace{-1mm}
\end{figure}

\paragraph{Newton-type methods.}

\begin{figure*}[t]
  \centering
    \includegraphics[width=1.0\linewidth]{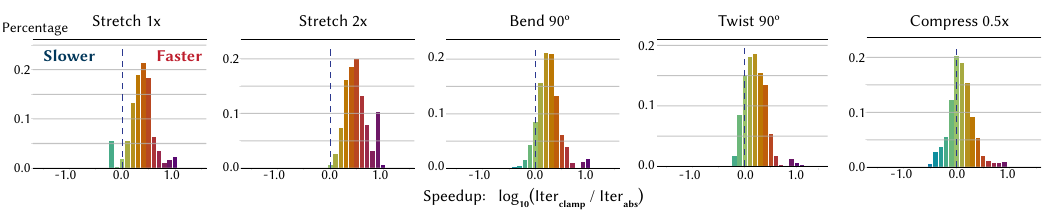}
    \caption{Histogram of the speedup of our abs strategy over the eigenvalue clamping strategy \cite{irving05} on the TetWild Thingi10k dataset \cite{Hu:2018:TMW:3197517.3201353, Thingi10K} with diverse deformations, including stretching (1x and 2x), compression, twisting, and bending.
    \changed{
      Our method offers significant improvements for large stretching, twisting and bending where previous methods may fail,
      while still, on average, being at least comparable in other regimes (e.g., compression).
    }
    }
    \label{fig:histogram_thingi10k}
    \vspace{-3mm}
\end{figure*}

Alternatively, there are Newton-type methods that do not rely on per-element PSD projection.
One strategy is to add a multiple of identity matrix to the local or global
hessian \cite{num_opt2006} but it tends to overdamp the convergence (see
\cite{liu2017quasi} Fig.8 and \cite{Shtengel2017majorization} Fig.2).
\citet{Shtengel2017majorization} proposed Composite Majorization, a tight
convex majorizer as an analytic PSD approximation of the Hessian. 
While this approximation is efficient to compute, it remains unclear how to extend it beyond 2D problems and to different types of energies.
Instead of directly approximating the Hessian matrix, \citet{Lan2023stencil}
proposed a strategy to adjust the searching direction derived
from the (possibly non-PSD) Hessian when performing stencil-wise Newton-CG. 
\citet{Chen14:ANM} applied the asymptotic numerical method to (inverse) static
equilibrium problem, and demonstrates its advantages over traditional
Newton-type methods.
When using incremental potential in dynamic settings,
\citet{longva2023pitfalls} proposed two alternative strategies: one is to use the original
Hessian matrix whenever it is positive definite and use an approximated Hessian
matrix otherwise; another is to add a multiple of the mass matrix to the original
Hessian until it becomes positive definite.
Both of these strategies could require one or several additional Cholesky
factorizations each Newton iteration anytime the original Hessian is not PSD.
Moreover, this method still inherits the flaws of those fallback Hessians
whenever they're invoked.
In contrast, our method maintains the same per-iteration computational cost as the
\emph{element-wise} projected Newton's method~\cite{irving05} while achieving a better
convergence rate compared to projected Newton \cite{irving05} and other
alternatives \cite{longva2023pitfalls}.

\section{Background}

\begin{figure}
  \centering
  \includegraphics[width=\linewidth]{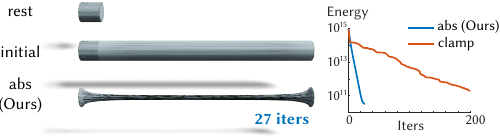}
  \caption{We stress \changed{test} our abs projection strategy under large stretching (11x) with high Poisson's ratio $\nu = 0.4999$.
  Here we create the initial deformation by moving the rightmost vertices of a cylinder by a factor of 10x.} 
  \label{fig:cylinder_large_stretch}
  \vspace{-1mm}
\end{figure}
Our approach focuses on quasi-static simulation, which amounts to solving a minimization problem for a (typically nonlinear) energy $f$ with respect to parameters $\vx$
\begin{align}
  \min_{\vx} f(\vx)\;.
\end{align}

Perhaps the most popular way of solving this problem is \changed{Newton's method}.
Given a set of parameters $\vx$ at the current iteration, Newton's method
approximates the energy $f$ locally with a quadratic function
\begin{align}\label{equ:taylor_approximation}
  \tilde{f}(\vx + \vd) \approx f(\vx) + \vg^T \vd + \frac{1}{2} \vd^T \mH \vd,
\end{align}
where $\vg = \nabla f(\vx)$ and $\changed{\mH} = \nabla^2 f(\vx)$ are the gradient and
the Hessian of the energy $f$ evaluated at $\vx$, and $\vd$ denotes the update
vector to the parameter $\vx$. 
To obtain the optimal $\vd$, one can set the derivative of $\nicefrac{\partial
    \tilde{f}}{\partial \vd}$ to zero and obtain the update with
\begin{align}
  \label{equ:step}
  \vd = - \mH^{-1} \vg.
\end{align}
Because $\tilde{f}$ is merely an approximation of the energy, the resulting $\vd$ may not guarantee
energy decrease at the end of the iteration.
Thus, a line search is needed to find a step size $\alpha$ such that $\vx + \alpha \vd$ achieves a sufficient decrease in the energy $f$.
Newton's method iterates between these steps until convergence.

\begin{wrapfigure}[16]{r}{1.1in}
  \vspace{-12pt}
  \includegraphics[width=\linewidth, trim={0mm 0mm 3mm 0mm}]{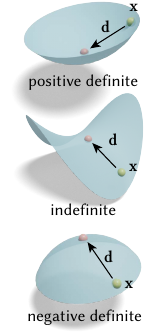}
  \label{fig:hessian_shapes}
\end{wrapfigure}
However, the process summarized above only works in the ideal situation where
the Hessian matrix $\mH$ is \emph{positive definite}. 
Geometrically, a positive definite Hessian $\mH$ corresponds to a convex energy space,
which ensures that the update direction $\vd$ towards the
\emph{critical point} $\nicefrac{\partial \tilde{f}}{\partial \vd} = 0$ is a
decent direction towards the minimum of $\tilde{f}$ (see inset). 
An indefinite and a negative definite Hessian $\mH$, however, correspond to a saddle and
a concave shape, respectively (see inset). In both cases, the direction
$\vd$ towards the critical point could be an energy \emph{ascent} direction, and thus
the Newton's method may fail to converge (see inset).
To address this issue, a popular approach is the (per-element) \emph{Projected Newton's method}.

\subsection{Projected Newton}\label{sec:projected_newton}

The idea of projected Newton's method is to approximate a non-positive definite Hessian matrix
using a positive definite one.
In the case of finite-element simulations, the (global) Hessian matrix $\mH_k$ can be assembled from the per-element Hessian matrix $\mH_i$ for each mesh (or tetrahedral) element $i$:
\begin{align}
  \mH = \sum_i \mP_i^\top \mH_i \mP_i,
\end{align}
where $\mP_i$ is a selection matrix that maps the per-element degrees of freedom to the global degrees of freedom.

The de facto way of making the global Hessian $\mH$ positive definite is to project the per-element Hessian $\mH_i$ to positive (semi-)definite (PSD) by performing eigen decomposition on the per-element Hessian $\mH_i$ numerically or analytically, followed by a clamping strategy to set the negative eigenvalues $\lambda_k$ to zero (or a small positive number $\epsilon$) \changed{\cite{irving05}}: 
\begin{align}\label{equ:clamping}
  \lambda_k^+ & = 
  \begin{cases}
    \epsilon & \text { if } \lambda_k \leq \epsilon, \\
    \lambda_k & \text { otherwise. }
    \end{cases}
\end{align}
Then, one can use the clamped eigenvalues $\lambda_k^+$ with the original
eigenvectors of $\mH_i$ of obtain a PSD projected per-element Hessian
$\mH^+_i$.
Although the projection happens locally, this guarantees the approximated global Hessian
\begin{align}
  \mH^+ = \sum_i \mP_i^\top \mH^+_i \mP_i
\end{align}
assembed from $\mH^+_i$ to be PSD \cite{Rockafellar1970convex}, thus a more robust simulation compared to using $\mH$.
This per-element Hessian modification strategy is also scalable because it only requires eigendecomposition on each (small) per-element Hessian $\mH_i$, instead of the global matrix.

\section{Pitfalls of Eigenvalue Clamping} \label{sec:pitfall_of_clamping}
\begin{wrapfigure}[7]{r}{1.1in}
  \vspace{-12pt}
  \includegraphics[width=\linewidth, trim={0mm 0mm 3mm 0mm}]{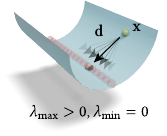}
  \label{fig:eigenvalue_clamp_shape}
\end{wrapfigure}
Although the projected Newton method has improved robustness over the vanilla
Newton's method, the widely used eigenvalue clamping strategy described in
\refsec{projected_newton} still suffers from poor convergence when the simulated
object undergoes large volume changes.
The cause lies in the operation in \refequ{clamping}, which
clamps the minimum eigenvalue $\lambda_\text{min}$ to zero (or a small positive number).
This operation effectively turns the local quadratic approximation from a
saddle shape (see inset in \refsec{projected_newton}) to a ``valley'' shape
(see inset). 
In this situation, the Newton update direction $\vd$ could point toward any
point on the bottom of the valley\textemdash including points that are far away from
the current parameter location $\vx$ and even points leading to energy increase.

\begin{figure}[t]
  \centering
  \includegraphics[width=\linewidth]{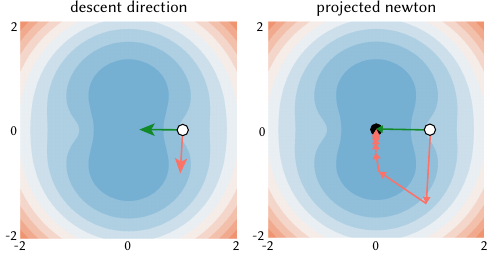} \caption{Consider a function $f(x)$, we visualize the descent direction (left) and the corresponding projected Newton trajectory (right). 
  Our abs-projection (green) takes 3 iterations to converge while $\epsilon$-projection direction (red) \cite{irving05}  takes 11 iterations.}
  \label{fig:newton_example_2_variable}
\end{figure}

To solidify this intuition, let us consider a minimal example which exemplifies the extreme failure case of eigenvalue clamping.
The function of two variables visualized in in
\reffig{newton_example_2_variable} is defined as 
\begin{equation}
    f(x, y) =
    \left( \sqrt{(x+1)^2+y^2} - 1 \right)^2 + \left( \sqrt{(x-1)^2+y^2} - 1
\right)^2. 
\end{equation}
At the point $(x,y) = (1-10^{-6}, 10^{-8})$ (indicated by the white point),
we have:
\begin{align}
  f = 2, \quad \nabla f=\begin{bmatrix} 3.99 \\ -0.02 \end{bmatrix}, \quad
  \mH 
  = \Phi
    \begin{bmatrix}
      -1.99\times10^6 & 0 \\
      0 & 3.99
    \end{bmatrix}
    \Phi^\top 
\end{align} 
with eigenvectors $\Phi = \begin{bmatrix}
  0.01 & -0.99 \\
  0.99 & 0.01
\end{bmatrix}$.

If we project the negative eigenvalues to a small positive value $\epsilon$
(e.g., $\delta = 10^{-3}$), the corresponding Newton direction can be computed
as
\begin{equation}  
  \begin{aligned} 
    -(\mH^+)^{-1} \nabla f
    = - \left(\Phi
    \begin{bmatrix}
      \epsilon & 0 \\
      0 & 3.99
    \end{bmatrix}
    \Phi^\top  \right)^{-1}
    \begin{bmatrix}
      3.99 \\
      -0.02
    \end{bmatrix} 
    = -\begin{bmatrix}
      1.19 \\
      19.99
    \end{bmatrix}.
  \end{aligned}
\end{equation}

For small $\epsilon$, the update direction is dominated by y-coordinate.
For extreme cases where $\epsilon = 10^{-9}$, the update direction even blows up along the y-coordinate.
Although $f$ technically deceases along the search direction, its projection violates the
spirit of Newton's method which relies on the quadratic approximation of the
energy function in the local region. 

The issue of eigenvalue clamping also appears in 3D. This is especially common
when simulating large deformations on materials with high Poisson ratios (see
\refsec{result}).

  In our 2D example, we can drive this bad eigenvalue 
  arbitrarily negative by moving the evaluation toward $(x,y)=(1,0)$.
  A more negative eigenvalue means the function along the corresponding eigendirection 
  is more concave, thus more poorly approximated by a convex quadratic.
  Unfortunately, clamping to near-zero effectively \emph{prefers} this direction
  (because after inversion the coefficient explodes, see \refequ{step}).
  We would rather like a filtering method which \emph{avoids} this direction.
  It is much more reasonable to make the effect of non-convex directions directly
  proportional to their eigenvalue magnitude.
  This immediately motivates using the absolute value as a filter.

\section{Absolute Value Eigenvalue Projection} \label{sec:abs_projection}

The analysis above suggests that projecting a large negative eigenvalue to a small positive value may lead to a poor estimate of the descent direction.
The optimization may get stuck in a local minimum or even diverge in this case. 
Inspired by \cite{gill1981practical} (Sec.4.4.2.1) and \cite{Paternain2019nonconvex_newton}, we propose to project the negative eigenvalues of the local Hessian $\mH_i$ to its absolute value:
\begin{align}
  \lambda_k^+ = \left| \lambda_k \right|.
\end{align}
\changed{The resulting projected Newton step can also be interpreted as the result of a generalized trust-region method where the model is a first-order Taylor expansion and the trust region is defined using the Hessian metric (see \cite{dauphin2014saddle_free}). }

\emph{Implementation.}
For implementations already using per-element projection and scatter-gather
assembly of Hessians following \citet{irving05}, our method is a single line
change (see Page~\pageref{code:change}).

For completeness, the full algorithm per-element absolute eigenvalue projection Hessian construction follows as:
\begin{minipage}[10pt]{0.95\linewidth}
  \begin{lstlisting}[language=Pseudocode,mathescape=true]
  initialize H_proj to empty sparse matrix
  foreach element i:
    # Pi*x selects element i's local variables from x
    either:
      Hi = constructLocalHessian(i,Pi*x)
      $\Lambda$,U = eig(Hi)
    or:
      $\Lambda$,U = analyticDecomposition(i,Pi*x)
    # Our one-line change
    Hi_proj = U*abs($\Lambda$)*U'
    # accumulate into global Hessian
    H_proj += Pi*Hi_proj*Pi'\end{lstlisting}
\end{minipage}

Applying our projection to the minimal example above, the Newton direction now becomes
\begin{equation}  
  \begin{aligned} 
    \vd & = -(\mH^+)^{-1} \vg 
    = - \left(\Phi
    \begin{bmatrix}
      \left| -1.99\times10^6 \right| & 0 \\
      0 & 3.99
    \end{bmatrix}
    \Phi^\top \right)^{-1}
    \begin{bmatrix}
      3.99 \\
      -0.02
    \end{bmatrix} \\
    & = - \left(\Phi
    \begin{bmatrix}
      5\times10^{-7} & 0 \\
      0 & 0.25
    \end{bmatrix} 
    \Phi^\top  \right)
    \begin{bmatrix}
      3.99 \\
      -0.02
    \end{bmatrix}
    = -\begin{bmatrix}
      0.99 \\
      -0.01
    \end{bmatrix}.
  \end{aligned}
\end{equation}
which is more aligned with the direction towards the global minimum (see \reffig{newton_example_2_variable}).

\begin{figure}
  \centering
  \includegraphics[width=\linewidth]{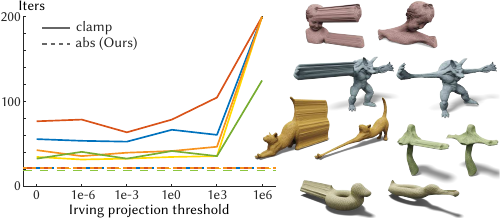}
  \caption{Our absolute projection strategy is parameter-free and achieves consistent speedup over the eigenvalue clamping strategy in the cases of high Poisson's ratio and large volume change. 
  On the contrary, for the eigenvalue clamping strategy, the optimal $\epsilon$ varies across different examples and picking the optimal one requires manual tuning.}
  \label{fig:different_eps}
\end{figure}

\section{Eigenanalysis of Volume-preserving Energy} \label{sec:snh_eigenanalysis}
While the two variable example may seem contrived, we show that the existence of
large negative eigenvalues is closely connected to the large Poisson's ratio and
large volume change due to the volume-preserving term in Neo-Hookean energy.

When the Hessian contains large negative eigenvalues, the absolute value
projection gives a better estimate of the descent direction, making the
optimization more stable and faster to converge. Then the question comes, when
does the Hessian contain large negative eigenvalues? It turns out that it
appears more often than we thought.

\begin{figure}[t]
  \includegraphics[width=\linewidth]{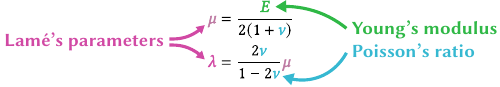}
  \caption{As Poisson's ratio $\nu \rightarrow \sfrac{1}{2}$, Lam\'e's second parameter $\lambda \rightarrow \infty$.}
  \label{fig:lame_youngs_poisson}
\end{figure}
The class of Neo-Hookean energy usually takes the form of 
\begin{align}
  \Psi = \frac{\mu}{2} (I_C - 3) - \mu \log(J) + \frac{\lambda}{2} (J-1)^2,
\end{align}
where $\mu$ and $\lambda$ are the first and second Lame parameters, $I_C = \text{tr}(\mF^\top \mF)$ is the first right Cauchy-Green invariant and $J = \text{det}(\mF)$ is the determinant of the deformation gradient $\mF$.

To ensure the inversion stability and rest stability, \changed{\citet{smith2018snh}} proposes to use the stable Neo-Hookean energy:
\begin{align}
  \Psi = \frac{\mu}{2}\left(I_C-3\right)+\frac{\lambda}{2}(J-\alpha)^2,
\end{align}
where $\alpha = 1+\frac{\mu}{\lambda}$.

As shown by the eigenanalysis in \changed{\cite{smith2018snh, kim22deformables}}, aside from the three eigenvalues corresponding to the scaling, the other six twist and flip eigenvalues of the local Hessian matrix for stable Neo-Hookean energy can be written as
\begin{align}
  & \Lambda_3=\mu+\sigma_z\left(\lambda\left(J-1\right)-\mu\right) \\
  & \Lambda_4=\mu+\sigma_x\left(\lambda\left(J-1\right)-\mu\right) \\
  & \Lambda_5=\mu+\sigma_y\left(\lambda\left(J-1\right)-\mu\right) \\
  & \Lambda_6=\mu-\sigma_z\left(\lambda\left(J-1\right)-\mu\right) \\
  & \Lambda_7=\mu-\sigma_x\left(\lambda\left(J-1\right)-\mu\right) \\
  & \Lambda_8=\mu-\sigma_y\left(\lambda\left(J-1\right)-\mu\right).
\end{align}
where $\mu$ and $\lambda$ are the Lame parameters, and $\sigma_x$, $\sigma_y$, and $\sigma_z$ are the singular \changed{values} of the deformation gradient $\mF$.

\begin{figure}[t]
  \centering
  \includegraphics[width=\linewidth]{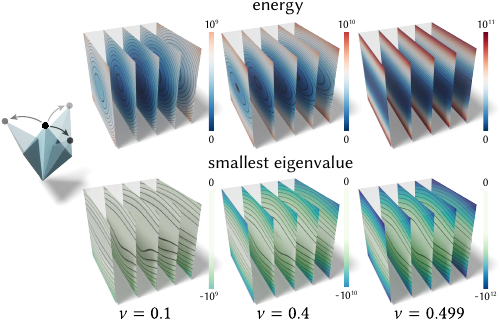} 
  \caption{Energy and smallest eigenvalues of stable Neo-Hookean energy with different Poisson's ratios $\nu$ and deformation.
  Here the deformation is created by dragging the top vertex of a regular tetrahedron around, and the center of the plot is with zero displacement.}
  \label{fig:snh_energy}
\end{figure}

When the Poisson's ratio $\nu$ is close to 0.5, the value of Lam\'e's second parameter $\lambda=\frac{2 \nu}{1-2 \nu} \mu$ is very large (see \reffig{lame_youngs_poisson}).
Thus the magnitude of potential negative eigenvalues $\Lambda_{6 \sim 8}$ largely depends on the Lam\'e's second parameter $\lambda$ and the volume change $(J-1)$.
For instance, a Poisson's ratio of 0.495 corresponds to a Lam\'e's second parameter of $\lambda \approx 100 \mu$.
This gives the eigenvalues $\Lambda_{6 \sim 8}$ a large negative value if there is a relatively large volume change for a specific element (i.e., when $(J -1)$ is large, see \reffig{snh_energy}). 

When the Hessian contains large negative eigenvalues, projecting it to a small positive value may lead to a poor estimate of the descent direction which strongly \changed{biases} towards the negative eigenvectors (see \refsec{pitfall_of_clamping}).
In contrast, the absolute value projection gives a better estimate of the descent direction, making the optimization more stable and faster to converge.

\section{Results}
\label{sec:result}
\begin{figure}
  \centering
  \includegraphics[width=\linewidth]{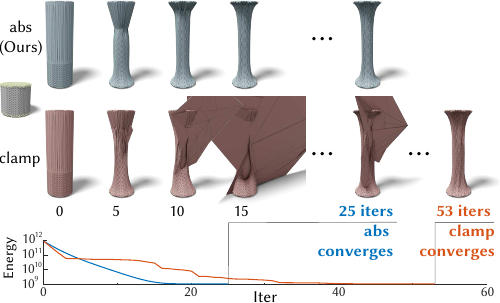}
  \includegraphics[width=\linewidth]{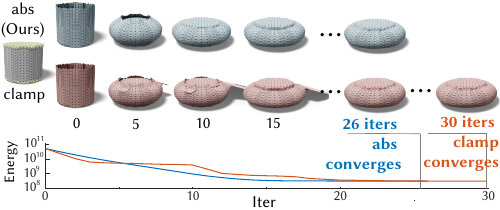}
  \caption{Our abs projection strategy stabilizes the optimization under large volume change and high Poisson's ratio,
  and achieves faster convergence rates than the traditional eigenvalue clamping strategy \cite{irving05}.
  Here we stretch a cylinder to 3.0x (top) and compress a cylinder to 0.5x (bottom) by moving the topmost vertices. } 
  \label{fig:cylinder_stretch_compress}
\end{figure}
  
We evaluate our method by comparing it against the traditional eigenvalue
clamping strategy \cite{irving05} and other alternatives on a wide range of
challenging examples, including different deformations, geometries, elastic
energies and physical parameters.
Furthermore, we experiment on the TetWild Thingi10k dataset \cite{Hu:2018:TMW:3197517.3201353, Thingi10K} with diverse deformations to 
test the robustness and scalability of our method.

Unless stated otherwise, we use stable Neo-Hookean model \changed{\cite{smith2018snh}} with Young's Modulus $E = 10^8$ and Poisson's ratio $\nu = 0.495$ for all the experiments, and use 0 as the threshold for the eigenvalue clamping strategy, as suggested by Sec.8 of \cite{irving05}.
\changed{We use a direct sparse Cholesky solver to solve the linear system at each Newton iteration.}
The convergence threshold is set to be when the newton decrement $0.5 \vd^\top \vg$ is less than $10^{-5}$ (times $\lambda$ for the scale of the gradient).
The initial deformation is created by moving some selected vertices of the mesh.
We run Newton's method for a 
maximum of 200 iterations with a 
\begin{wrapfigure}[5]{r}{1.4in}
  \vspace{-15pt}
  \includegraphics[width=\linewidth, trim={0mm 0mm 3mm 0mm}]{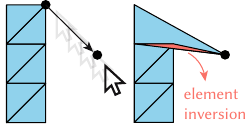}
  \label{fig:2d_inversion}
\end{wrapfigure}
classical backtracking line search strategy \cite{num_opt2006}.
As inversion is extremely common in the presence of large initial deformation (see the inset), we do not use an inversion-aware line search.

We implement our method in C++ with libigl \cite{libigl} and use TinyAD \cite{schmidt2022tinyad} for the automatic differentiation.
Experiments are performed using a MacBook Pro with an Apple M2 processor and 24GB of RAM.

\begin{figure}[t]
  \centering
  \includegraphics[width=\linewidth]{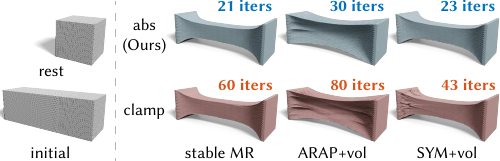}
  \caption{Our abs projection strategy generalizes over various volume-preserving hyperelastic models. 
  Here we add the volume term $(J-1)^2$ to different strain energies, including Mooney-Rivlin \cite{smith2018snh}, ARAP \cite{lin2022isoARAP} and Symmetric Dirichlet \cite{bijective_parameterization} energy.}
  \label{fig:different_energy}
\end{figure}
\begin{figure}[t]
  \centering
  \includegraphics[width=\linewidth]{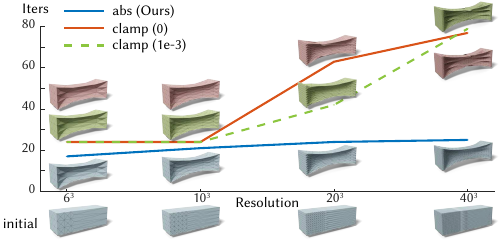}
  \caption{The speedup of our abs projection strategy over the eigenvalue clamping strategy increases with the mesh resolution.
  Here we stretch the rightmost vertices of a cube (of 4 different resolutions) by a factor of 3.0x. }
  \label{fig:different_resolution}
\end{figure}
\begin{figure}
  \centering
  \includegraphics[width=\linewidth]{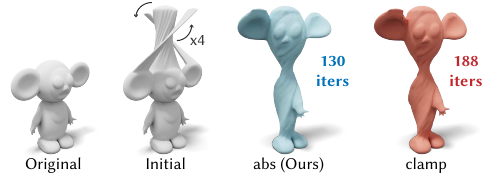}
  \caption{\changed{
  Our method stabilizes and accelerates the optimization in the presence of large rotations.
  Here we stretch and twist the topmost vertices of the doll by 360 degrees (divided into four 90° subsolves to avoid ambiguity). }
  }
  \label{fig:doll_twist_360}
\end{figure}
\begin{figure}[t]
  \centering
  \includegraphics[width=\linewidth]{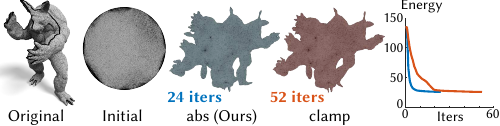}
  \caption{Our method can also accelerate the optimization of volume-preserving parameterization, where we minimize the energy $0.5 E_{\text{MIPS}} + 0.5 (J-1)^2$.
  Here we start from the Tutte embedding with the same total area as the original surface mesh (the cut is visualized in black), but nevertheless the per-element area distortion could still be large. } 
  \label{fig:parameterization}
\end{figure}
\begin{figure}
  \centering
  \includegraphics[width=\linewidth]{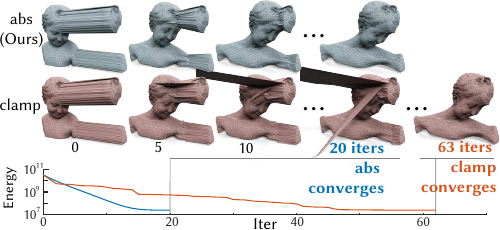}
  \includegraphics[width=\linewidth]{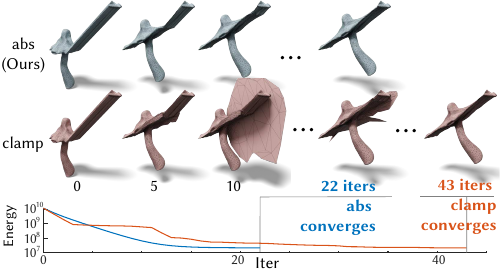}
  \includegraphics[width=\linewidth]{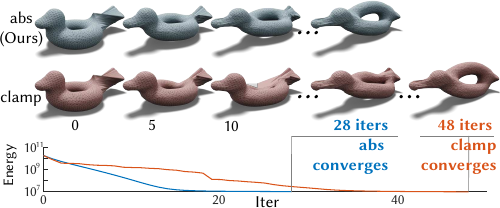}
  \caption{Our abs projection strategy enables stability and acceleration over diverse deformations with large volume change, including stretching (top), shearing (middle) and twisting (bottom).} 
  \label{fig:stretch_shear_twist}
\end{figure}
\begin{figure}
  \centering
  \includegraphics[width=\linewidth]{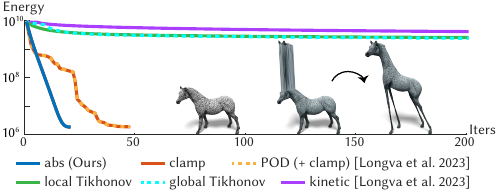}
  \caption{We compare our abs projection strategy with eigenvalue clamping strategy \cite{irving05}, adding a multiple of Identity matrix to the local Hessian or the global Hessian (until it becomes PSD) \cite{martin2011example}, and the Projection-on-Demand and Kinetic strategy \cite{longva2023pitfalls}.
  Note that the Projection-on-Demand strategy \cite{longva2023pitfalls} requires an additional Cholesky factorization to check the positive-definiteness of the original Hessian at each Newton iteration.
  }
  \label{fig:comparison_longva}
\end{figure}
\begin{figure}
  \centering
  \includegraphics[width=\linewidth]{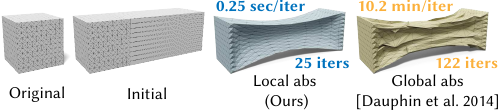}
  \caption{\changed{
  Compared to our local approach, the global abs approach \cite{dauphin2014saddle_free,Paternain2019nonconvex_newton} is computationally intractable due to a full eigendecomposition of the global Hessian (takes more than 3 hours for one $38k \times 38k$ Hessian in \reffig{comparison_longva}), and leads to a damped convergence and suboptimal configurations (right).
  Here on average, the global abs approach takes 10.2 minutes per Newton iteration for one $8.8k \times 8.8k$ Hessian, while our local abs approach takes only 0.25 seconds for one Newton iteration.
  }}
  \label{fig:global_proj}
\end{figure}

\emph{Convergence and Stability.} We compare our abs projection strategy to \cite{irving05} under a variety of deformations, including stretching, compression, shearing and twisting in \reffig{stretch_shear_twist}, \reffig{cylinder_stretch_compress}, \reffig{armadillo_stretch}, \reffig{doll_twist_inv_ls} \changed{and \reffig{doll_twist_360}}.
Our abs projection strategy converges swiftly and smoothly under large local volume change and high Poisson's ratio, while \cite{irving05} suffers from a much slower convergence rate due to the instability in the optimization.
\changed{The energies and shapes for both methods are generally consistent within each example when converged, except in a few cases where the other method converges to a different local minimum after “blowing up” mid-optimization (e.g., \reffig{teaser} (clamp) and \reffig{global_proj} (global abs)). }
\changed{We also compare our local abs approach to the global abs approach \cite{dauphin2014saddle_free,Paternain2019nonconvex_newton} in \reffig{global_proj}.
Compared to our method, the global abs approach is computationally intractable due to a full eigendecomposition of the global Hessian (takes more than 3 hours for one $38k \times 38k$ Hessian in \reffig{comparison_longva}), 
and leads to a damped convergence and suboptimal configurations (\reffig{global_proj}).}
We further stress test our method under extreme volume change (stretch to 11x) and an even higher Poisson's ratio ($\nu = 0.4999$) in \reffig{cylinder_large_stretch}.
Our method still maintains a stable and fast convergence rate, while \cite{irving05} still fails to converge after 200 iterations.

\emph{Generalization.} We further evaluate the generality of our method over different mesh resolutions, Poisson's ratios, mesh deformations and hyperelastic models. 
Under the same deformation, the speedup we gain over \cite{irving05} increases with the mesh resolutions (\reffig{different_resolution}).
Our abs projection approach maintains a relatively stable convergence rate across different resolutions while the traditional eigenvalue clamping method requires drastically more iterations as the resolution increases.
In \reffig{different_pr}, we show the Newton iteration counts of our method and \cite{irving05} under different Poisson's ratios and volume changes.
For relatively small volume change (top), the speedup of our method over \cite{irving05} appears after, e.g., the Poisson's ratio is larger than 0.47.
But for volume change that is large enough (bottom), our method can accelerate the convergence rate even when the Poisson's ratio is low, e.g., $\nu = 0.3$.
We further demonstrate the growth of our speedup as the initial volume change increases in \reffig{different_deformation_size}. 
Our abs projection strategy can also generalize to other tasks, such as surface parameterization in \reffig{parameterization}, \changed{stable Neo-Hookean energy with collisions in \reffig{collision_horse},} and other volume-preserving hyperelastic models in \reffig{different_energy}, including Mooney-Rivlin \cite{smith2018snh}, ARAP \cite{lin2022isoARAP} and Symmetric Dirichlet energy (each with an additional volume term $(J-1)^2$ to make it volume-preserving).

\begin{figure}
  \centering
  \includegraphics[width=\linewidth]{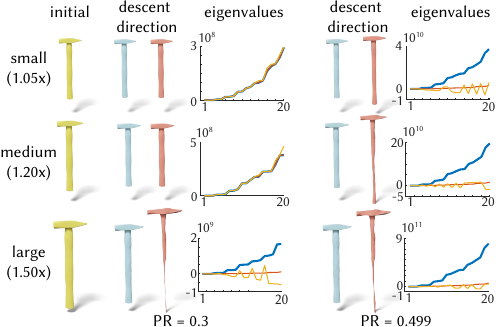}
  \caption{We visualize the first 20 eigenvalues of the Hessian and the resulting descent direction using our abs projection strategy (red) and the eigenvalue clamping strategy (blue) under different volume changes and Poisson's ratio.
  Here we uniformly scale the mesh by 1.05x, 1.2x and 1.5x. 
  The eigenvalues of the original Hessian are denoted by yellow. }
  \label{fig:eigenmodes_volume}
\end{figure}
\emph{Comparison. } 
We evaluate our method and \cite{irving05} (with threshold $10^{-3}$) on a total of 593 the closed, genus-0 high-resolution tetrahedral meshes (with more than 5,000 vertices) from the TetWild Thingi10k dataset \cite{Hu:2018:TMW:3197517.3201353, Thingi10K} with diverse deformations, including stretching, stretching (2x), compression, twisting and bending.
As shown by the histogram in \reffig{histogram_thingi10k}, our method is at least comparable with the eigenvalue clamping and offers, on average, 2.5 times speedup for large deformations.
In \reffig{different_eps}, since the eigenvalue clamping strategy contains an additional user-picked parameter $\epsilon$ to control the clamping threshold, we compare the convergence rate of our method and \cite{irving05} with different $\epsilon$.
Our method is parameter-free and achieves consistent speedup over \cite{irving05}.
We further compare our abs projection scheme with other alternatives in \reffig{comparison_longva}, including adding a multiple of Identity matrix to the local Hessian or the global Hessian (until it becomes PSD) \cite{martin2011example}, and the Projection-on-Demand and Kinetic strategy \cite{longva2023pitfalls}.
Adding a multiple of Identity or mass matrix to the Hessian eventually makes the optimization closer to (preconditioned) gradient descent, and thus damps the convergence too much for large deformation problems.
Note that the Projection-on-Demand strategy \cite{longva2023pitfalls} requires an additional Cholesky factorization to check the positive-definiteness of the original Hessian at each Newton iteration.

\section{CONCLUSION \& FUTURE WORK}
\begin{figure}
  \centering
  \includegraphics[width=\linewidth]{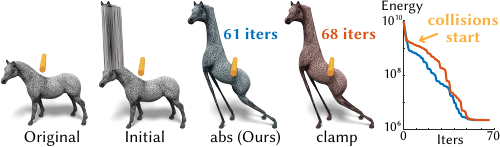}
  \caption{\changed{
  We perform a collision experiment using Incremental Potential Contact (IPC) \cite{Li2020IPC}, where we put a cylinder (orange) above the back of a horse.
  Our method is still able to achieve speedup with collisions, even though IPC's intersection-aware line search clamps down the step size and dominates convergence after collisions happen.
  }}
  \label{fig:collision_horse}
\end{figure}
We propose a Hessian modification strategy to improve the robustness and efficiency of hyperelastic simulations. 
\begin{wrapfigure}[9]{r}{1.1in}
  \vspace{-12pt}
  \includegraphics[width=\linewidth, trim={0mm 0mm 3mm 0mm}]{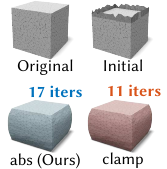}
  \label{fig:limitation_cube}
\end{wrapfigure}
Our method is a simple one-line change to the existing projected Newton's method, making it extremely reproducible and widely applicable in many simulation frameworks. 
We primarily evaluate our method on Neo-Hookean simulations with large deformations. 
For small deformation, especially with compression (see the inset), our method can sometimes slightly damp the convergence compared to \cite{irving05}.
Extending our analysis to a wider range of finite element simulations \changed{(e.g., collisions (\reffig{collision_horse}))} could better identify the applicability of our approach, and potentially deriving an even better adaptive PSD projection method. 
Exploring other Hessian modification strategies, such as adding higher-order regularization terms to the elastic energy \cite{kim22deformables} or a combination with \cite{longva2023pitfalls}, could also be an interesting future direction.
Considering using another volume-preserving term that does not introduce large negative eigenvalues in the presence of large deformations could \changed{be another promising} future direction.

\vspace{-2pt}
\begin{acks}
This work is funded in part by two NSERC Discovery grants, the Ontario Early Research Award program, the Canada Research Chairs Program, a Sloan Research Fellowship, the DSI Catalyst Grant program, SoftBank and gifts by Adobe Research and Autodesk.
We thank Danny Kaufman for early discussions;
Yuta Noma for testing the code; 
all the artists for sharing the 3D models and anonymous reviewers for their helpful comments and suggestions.
\end{acks}

\bibliographystyle{ACM-Reference-Format}
\bibliography{sections/reference.bib}


\end{document}